\documentclass[letter,times,twocolumn,tighten,trackchanges]{aastex62}
\hypersetup{linkcolor=red,citecolor=blue,filecolor=cyan,urlcolor=blue}
\usepackage{amsmath,amstext}
\usepackage[figure,figure*]{hypcap}
\usepackage{commath}

\definecolor{oxfordblue}{rgb}{0.0, 0.13, 0.28}
\definecolor{firebrick1}{rgb}{0.698, 0.133, 0.133}

\shorttitle{Dust continuum and gas kinematics of HD\,100546}
\shortauthors{P\'erez et al.}

\begin{document}

\title{\bf \large Long Baseline Observations of the HD\,100546 Protoplanetary Disk with ALMA}



\correspondingauthor{Sebasti\'an P\'erez}
\email{sebastian.astrophysics@gmail.com}
\author[0000-0003-2953-755X]{Sebasti\'an P\'erez}

\affiliation{Universidad de Santiago de Chile, Av. Libertador Bernardo O'Higgins 3363, Estaci\'on Central, Santiago, Chile}
\affiliation{Departamento de Astronom\'ia, Universidad de Chile, Casilla 36-D, Santiago}

\author[0000-0002-0433-9840]{Simon Casassus}
\affiliation{Departamento de Astronom\'ia, Universidad de Chile, Casilla 36-D, Santiago}

\author[0000-0001-5073-2849]{Antonio Hales}
\affiliation{National Radio Astronomy Observatory, 520 Edgemont Road, Charlottesville, VA 22903-2475, USA}
\affiliation{Atacama Large Millimeter/Submillimeter Array, Joint ALMA Observatory, Alonso de C\'ordova 3107, Vitacura 763-0355, Santiago, Chile}

\author[0000-0002-5352-2924]{Sebasti\'an Marino}
\affiliation{Max Planck Institute for Astronomy (MPIA), K\"onigstuhl 17, 69117 Heidelberg, Germany}

\author[0000-0003-3943-4044]{Anthony Cheetham}
\affiliation{Observatoire de Gen\`eve, Universit\'e de Gen\`eve, 51 chemin des Maillettes, 1290, Versoix, Switzerland}
\affiliation{Max Planck Institute for Astronomy (MPIA), K\"onigstuhl 17, 69117 Heidelberg, Germany}

\author[0000-0002-5903-8316]{Alice Zurlo}
\affiliation{Universidad Diego Portales, Av. Ejercito 441, Santiago, Chile}
\affiliation{Escuela de Ingenier\'ia Industrial, Facultad de Ingenier\'ia y Ciencias, Universidad Diego Portales, Av. Ejercito 441, Santiago, Chile}

\author[0000-0002-2828-1153]{Lucas Cieza}
\affiliation{Universidad Diego Portales, Av. Ejercito 441, Santiago, Chile}

\author[0000-0001-9290-7846]{Ruobing Dong}
\affiliation{Department of Physics \& Astronomy, University of Victoria, Victoria, BC, V8P 1A1}

\author[0000-0002-2692-7862]{Felipe Alarc\'on}
\affiliation{Department of Astronomy, University of Michigan, 1085 S. University, Ann Arbor, MI 48109, USA}

\author[0000-0002-3728-3329]{Pablo Ben\'itez-Llambay}
\affiliation{Niels Bohr International Academy, Niels Bohr Institute, Blegdamsvej 17, DK-2100 Copenhagen \O{}, Denmark}

\author[0000-0002-9036-2747]{Ed Fomalont}
\affiliation{National Radio Astronomy Observatory, 520 Edgemont Road, Charlottesville, VA 22903-2475, USA}
\affiliation{Atacama Large Millimeter/Submillimeter Array, Joint ALMA Observatory, Alonso de C\'ordova 3107, Vitacura 763-0355, Santiago, Chile}

\author[0000-0002-1302-4613]{Henning Avenhaus}
\affiliation{Max Planck Institute for Astronomy (MPIA), K\"onigstuhl 17, 69117 Heidelberg, Germany}

\begin{abstract}                
  Using the Atacama Large Millimeter/submillimeter Array (ALMA), we
  observed the young Herbig star HD\,100546, host to a prominent disk
  with a deep, wide gap in the dust. The high-resolution 1.3 mm
  continuum observation reveals fine radial and azimuthal
  substructures in the form of a complex maze of ridges and trenches
  sculpting a dust ring. The $^{12}$CO(2--1) channel maps are
  modulated by wiggles or kinks that deviate from Keplerian kinematics
  particularly over the continuum ring, where deviations span
  90$^\circ$ in azimuth, covering $\sim$5~km~s$^{-1}$. The most
  pronounced wiggle resembles the imprint of an embedded massive
  planet of at least 5~M$_{\rm Jup}$ predicted from previous
  hydrodynamical simulations \citep*{Perez2018}. Such planet is
  expected to open a deep gap in both gas and dust density fields
  within a few orbital timescales, yet the kinematic wiggles lie near
  ridges in the continuum. The lesser strength of the wiggles in the
  $^{13}$CO and C$^{18}$O isotopologues show that the kinematic
  signature weakens at lower disk heights, and suggests qualitatively
  that it is due to vertical flows in the disk surface. Within the
  gap, the velocity field transitions from Keplerian to strongly
  non-Keplerian via a twist in position angle, suggesting the presence
  of another perturber and/or an inner warp. We also present
  VLT/SPHERE sparse aperture masking data which recovers scattered
  light emission from the gap's edges but shows no evidence for signal
  within the gap, discarding a stellar binary origin for its opening.
\end{abstract}



\section{Introduction} \label{sec:intro}

Substructures in protoplanetary disks are not only ubiquitous but also
diverse. Radial gaps, fine and wide rings, spiral arms, and lopsided
concentrations are amongst the most common substructures found in
resolved observations of disks. The mechanisms behind the formation of
such varied structures are not entirely known, but most models invoke
planet-disk interactions. It appears that these substructures are a
common feature of the early evolution of protoplanetary disks and it
is widely thought that they are related to the process of planet
formation.

Radial discontinuities and concentric rings appear to be the most
frequent substructures, such as in HL\,Tau \citep{ALMA2015}, TW\,Hya
\citep{Andrews2016}, HD\,97048 \citep{vanderPlas2017}, HD\,169142
\citep{Fedele2017, Perez2019}, and the disks in the Disk Substructures
at High Angular Resolution Project
\citep[DSHARP,][]{Andrews2018}. These dusty annuli can be sculpted by
planets \citep[e.g.,][]{Dipierro2015} which usually leave a clear gap
in the dust radial profile.  In the case of a low-mass planet, the
planet-disk interaction can produce multiple narrow gaps and rings
\citep[e.g.,][]{Dong2017}. For example, \citet{Perez2019} showed that
a migrating mini-Neptune planet can reproduce the three fine rings
observed in the HD169142 transition disk \citep[see
  also][]{Weber2019}.

Simulations of planet-disk interactions show that young massive
planets develop a {\em circumplanetary disk} (CPD) as they accrete
material from their parent protoplanetary disk
\citep[e.g.,][]{Miki1982, Lubow1999, Gressel2013, Szulagyi2014}. The
CPDs persist for as long as the planet grows. CPDs, however, are
difficult to detect in dust continuum \citep{Wu2017, Ricci2017,
  PerezMarino2019}. The protoplanet candidates inside the gap of
PDS\,70 are the only detections that have not been challenged so far
\citep{Keppler2018, Christiaens2019}. Recently, sub-millimeter
emission attributed to dust emission from a CPD around PDS70c has been
reported, as well as dust orbiting in proximity of PDS\,70b
\citep{Isella2019}. Indeed, the radio flux emitted by these CPDs may
be very faint because millimeter dust is believed to be lost by radial
drift within a few hundred years \citep{ZhuIsella2018}.

In \citet{Perez2015}, we show that a giant planet embedded in a
circumstellar disk produces distinct kinematical signatures,
detectable in CO velocity maps when probed at high resolution and
sensitivity. One of the predicted characteristics of a massive planet
is a wiggle or kink in the iso-velocity contours of CO emission in the
vicinity of the planet. The terms {\em twist} \citep{Perez2015}, {\em
  kink} \citep{Pinte2018} and {\em wiggle} \citep*{Perez2018} have all
been used to refer to a kinematic deviation associated with an
embedded planet. In this work, we opt for the word {\em
  wiggle}. \citet{Pinte2018} observed such a wiggle in iso-velocity
maps of HD\,163296, identifying the presence of a giant planet at
260~au. \citet{Teague2018}, using azimuthally-averaged kinematic
information, measured the pressure profile of a gap that is consistent
with a planet opening mechanism.  Recently, \citet{Pinte2019} also
found a wiggle coincident with the gap in the continuum map of
HD\,97048.

In this Letter, we present 1.3 mm observations of HD\,100546 at
$\lesssim$2 au resolution in continuum and $\lesssim$8 au resolution
in the CO(2--1) isotopologues molecular lines
(Section~\ref{sec:obs}). We describe the substructures observed in the
continuum in Sec.\,\ref{sec:continuum}. The $^{12}$CO gas moment maps
are described in Sec.\,\ref{sec:kinematics}. A kinematic wiggle
coincident with the continuum ring is presented in
Sec.\,\ref{sec:wiggle}. The deviations from the azimuthally symmetric
flow are quantified and discussed in a companion Letter
\citep{Casassus2019}. The velocity perturbations where the wiggle is
more pronounced show similarities with those expected for an accreting
giant. New infrared observations, presented in Sec.\,\ref{sec:sam},
rule out the presence of stellar companions inside the dust gap. A
discussion of the nature of the kinematic detection and the puzzling
association with the bright continuum ring is presented in Section
\ref{sec:discussion}. Throughout this Letter, we assume a distance of
110.0$\pm$0.6~pc to HD\,100546 \citep[][]{Gaia2018}.

\section{Observations} \label{sec:obs}

\subsection{ALMA continuum and molecular line observations}

We obtained 1.3 millimeter observations of HD\,100546 with the Atacama
Large Millimeter/submillimeter Array (ALMA) by combining the 12-m
array in extended (C40-9) and compact (C40-6) configurations, in the
context of Cycle\,4 project 2016.1.00344.S. The resulting baselines
ranged from 19~meters to up to 12.2~kilometers with a total of 39-42
antennas. The combined observations are sensitive to spatial scales of
up to 1\farcs5. The long baseline observations were acquired on
September 19, 22, and 23, in three different blocks of $\sim$90~min
each (40~min on source). Precipitable water vapor ranged between 0.3
and 0.8~mm.  Observations of atmospheric (J1147-6753), bandpass
(J0635-7516), and flux (J1107-4449) calibrators were performed. The
phase calibrator (J1147-6753) was alternated with the science target
to calibrate the time-dependent variation of the complex gains. The
cycling time for phase calibration was set to 8~minutes and 54~seconds
for the compact and extended configurations, respectively. The ALMA
correlator was configured in Frequency Division Mode (FDM). Two
spectral windows with 1.875~GHz bandwidth were set up for detecting
the dust continuum, centered at 232.005~GHz and 218.505~GHz,
respectively. The $^{12}$CO(2--1), $^{13}$CO(2--1) and C$^{18}$O(2--1)
transitions of carbon monoxide were targeted by configuring three
spectral windows at rest frequencies of 230.538~GHz, 220.399~GHz and
219.560~GHz respectively. The spectral resolution for the line
observations was 122.070~kHz (equivalent to 0.2~km~s$^{-1}$ channels).

All data were calibrated by the ALMA staff using the ALMA Pipeline
version 40896 in the CASA package \citep{McMullin2007}, including
offline Water Vapor Radiometer (WVR) calibration, system temperature
correction, as well as bandpass, phase, and amplitude
calibrations. The short baseline and long baseline datasets were
calibrated independently.

Synthesis imaging was initially carried out using the CLEAN algorithm
(CASA version 5.4, task {\tt tclean}). Self-calibration of the data
was performed to improve coherence. An SNR of $\sim$80 was achieved
prior to self-calibration. One round of phase self-calibration using a
solution interval of 54s was applied to the data, which was found to
improve the resulting SNR by a factor of 1.6.  A positional offset
between short and long baseline was corrected prior to combining the
datasets. An image reconstructed using natural weights, after
self-calibration and concatenation of the datasets, yields an RMS
noise of 12$\mu$Jy~beam$^{-1}$, for a CLEAN beam of 64$\times$45~mas.

As HD\,100546 is bright in the mm, we super-resolved the
self-calibrated continuum data using non-parametric image modeling
with the {\tt uvmem} package \citep[here we used the
  publicly-available GPU adaptation {\tt gpuvmem}\footnote{{\tt
      https://github.com/miguelcarcamov/gpuvmem}.},][]{Carcamo2018}. Image
positivity provided enough regularization (i.e. we did not add entropy
to the objective function). The {\tt uvmem} reconstruction
$I_{1.3\mathrm{mm}}$ provides slightly higher angular resolutions than
super-uniform weighting, but without compromising sensitivity
\citep[see][for details]{Carcamo2018}. We adopted the {\tt uvmem}
image for our analysis.

The effective angular resolution of the {\tt uvmem} model was measured
with an elliptical Gaussian fit to a deconvolution of a simulation
with the same $uv$-coverage as the ALMA observation taken on a single
spike. The input spike flux was 10~mJy, which is comparable with the
emission at the location of the central star. The final
$I_{1.3\mathrm{mm}}$ image has an angular resolution of $\Omega_b =
19.7\times12.9$~mas, or 2.1$\times$1.4~au.  This effective beam is
consistent with the expectation of $\lesssim 1/3$ the natural beams
\citep[][]{Carcamo2018}, which is $\Omega_\mathrm{nat} = 64\times$45
mas. We oversampled $\Omega_b$ with 2.5 mas pixels. The peak in flux
is 0.92 mJy\,beam$^{-1}$, at the location of the star.  The flux
density of the central source is 3.1$\pm$0.3\,mJy, while the total
flux density over the entire image is 436$\pm$40~mJy.

We estimate that the noise level in the $I_{1.3\mathrm{mm}}$ image is
16.6~$\mu$Jy\,beam$^{-1}$, where beam = $19.7\times12.9$~mas. The
following conservative approach was used to obtain this estimate.  We
first measured the noise in the restored image (comparable to a CLEAN
restoration), in natural weights. This was done by calculating the
standard deviation inside small boxes devoid of bright continuum
emission, but including synthesis imaging artifacts. This measurement
was repeated over several regions, and the largest standard deviation
was adopted as systematic noise. This amounts to $\sigma_{\rm
  total}\!=\!56$~$\mu$Jy~beam$^{-1}$, in the $\Omega_\mathrm{nat}$
beam.  Second, we assummed that the noise in $I_{1.3\mathrm{mm}}$ was
worse by a factor of $\sqrt{N_b}$, where $N_b \sim 11.4 =
\Omega_\mathrm{nat}/\Omega_b$ is the number of {\tt uvmem} beams
inside the natural-weight beams. The resulting noise level in
$I_{1.3\mathrm{mm}}$ is thus $\sigma_{\rm
  MEM}\!=\!16.6~\mu$Jy\,beam$^{-1}$, in the $\Omega_b$ {\tt uvmem}
beam, after dividing for $\Omega_\mathrm{nat}/\Omega_b$. A similar
estimate of the noise level can be obtained by assuming that a perfect
deconvolution would yield the same point-source sensitivity as the
dirty map in natural weights, with the thermal noise in natural
weights.

Channel maps of CO(2--1) emission were constructed with {\tt
  tclean}. We used a Briggs weighting scheme with a robust parameter
of 1.0 for $^{12}$CO, which yield the best results in terms of
achieving good signal-to-noise without compromising on
resolution. Channel maps were produced with a spectral resolution of
0.5 km~s$^{-1}$. Each channel map has an RMS noise of 1.3
mJy~beam$^{-1}$, for a CLEAN beam of 76$\times$57~mas. Line emission
cubes were produced with and without prior continuum-subtraction
(performed with CASA task {\tt uvcontsub}). For the $^{13}$CO and
C$^{18}$O isotopologues, a Briggs robust parameter of 2.0 was used
given the low signal in those lines, yielding maps with a beam
resolution of 82$\times$61~mas and RMS noise of 1.8 and 1.2
mJy~beam$^{-1}$, respectively. Here, we present the CO maps with
continuum subtraction \citep[$^{12}$CO maps without prior continuum
  subtraction can be found in ][]{Casassus2019}.  The $^{12}$CO moment
zero was calculated using an intensity-weighted sum along the spectral
axis, while first and second moments were determined via Gaussian fits
to the velocity profile along each pixel.  In the case of the cubes
without prior continuum-subtraction, the continuum emission is
accounted by a polynomial baseline when producing moment maps from
Gaussian fits. The Gaussian fits work best at recovering the broad and
velocity-structured emission profiles within the gap. The RMS noise in
the moment zero map is $\sim$3 mJy~beam$^{-1}$~km~s$^{-1}$. We checked
that channelization effects \citep[e.g.][]{Christiaens2014} are not
present in the moment maps by producing a cube with smaller channels
(0.25 km~s$^{-1}$) shifted by 0.1 km~s$^{-1}$ in velocity. The low
signal in $^{13}$CO and C$^{18}$O prevent the recovery of good quality
moment maps.

\subsection{SPHERE sparse aperture masking}

To assess whether there are any stellar companions within the gap , we
performed sparse aperture masking (SAM) observations with SPHERE IRDIS
and IFS instruments on VLT. The observations were acquired in May
15-16 2018, in the K1K2 band, with 1h integration on HD\,100546 plus
1h on calibrators. The data processing follows the same procedure as
in \citet{Cheetham2019}. The data were cleaned using the SPHERE Data
Reduction and Handling (DRH) pipeline \citep{Pavlov2008}, including
background subtraction, flat fielding and extraction of the spectral
data cube. IRDIS and IFS images are produced using the MIRA package
\citep{Thibaut2008} applied to the closure phases and visibilities,
with a hyperbolic regularization term. The weights in the
regularization were varied in order to suppress speckles.

\begin{figure*}[t]
  \centering
  \includegraphics[width=0.37\textwidth]{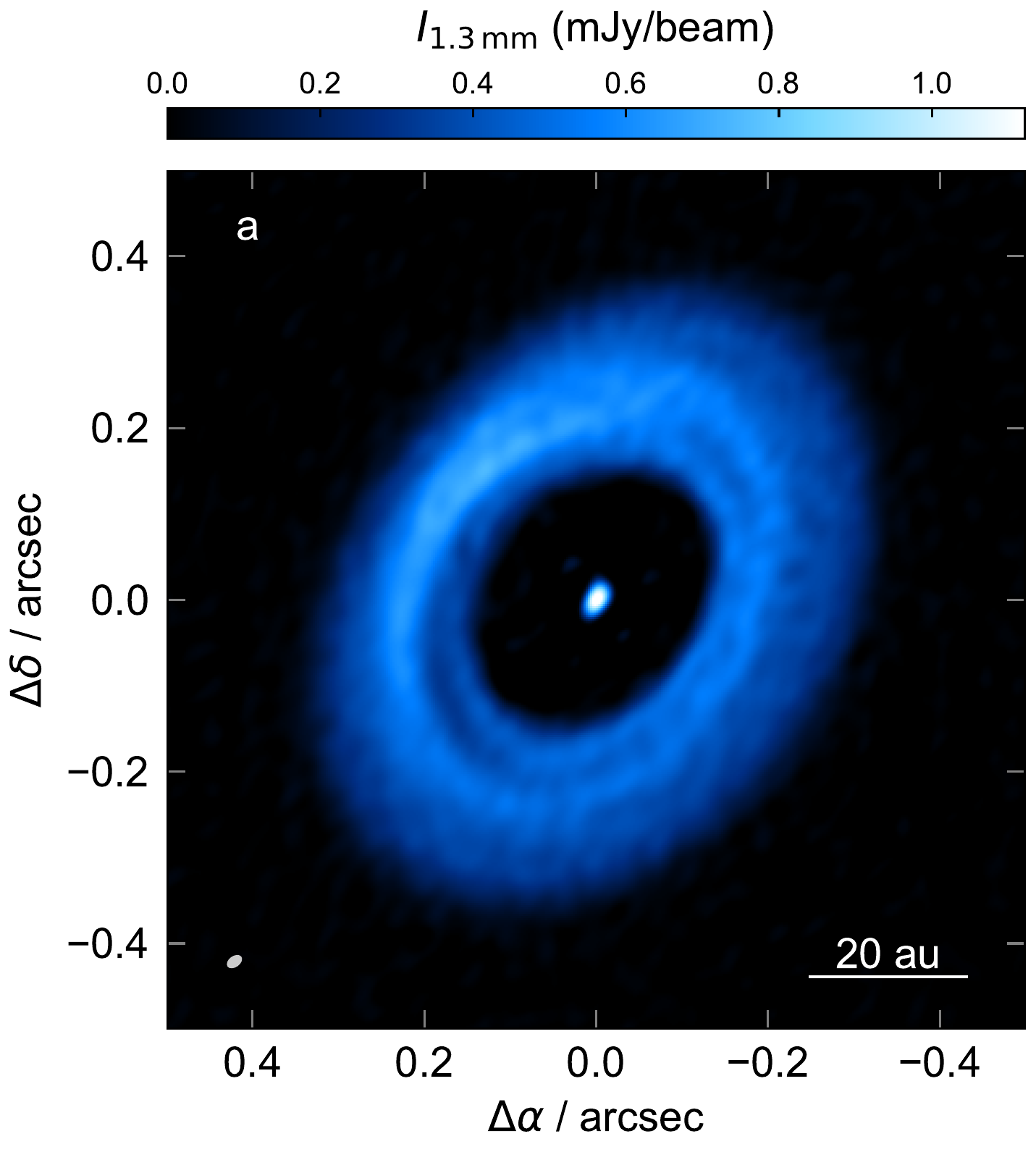}
  \includegraphics[width=0.37\textwidth]{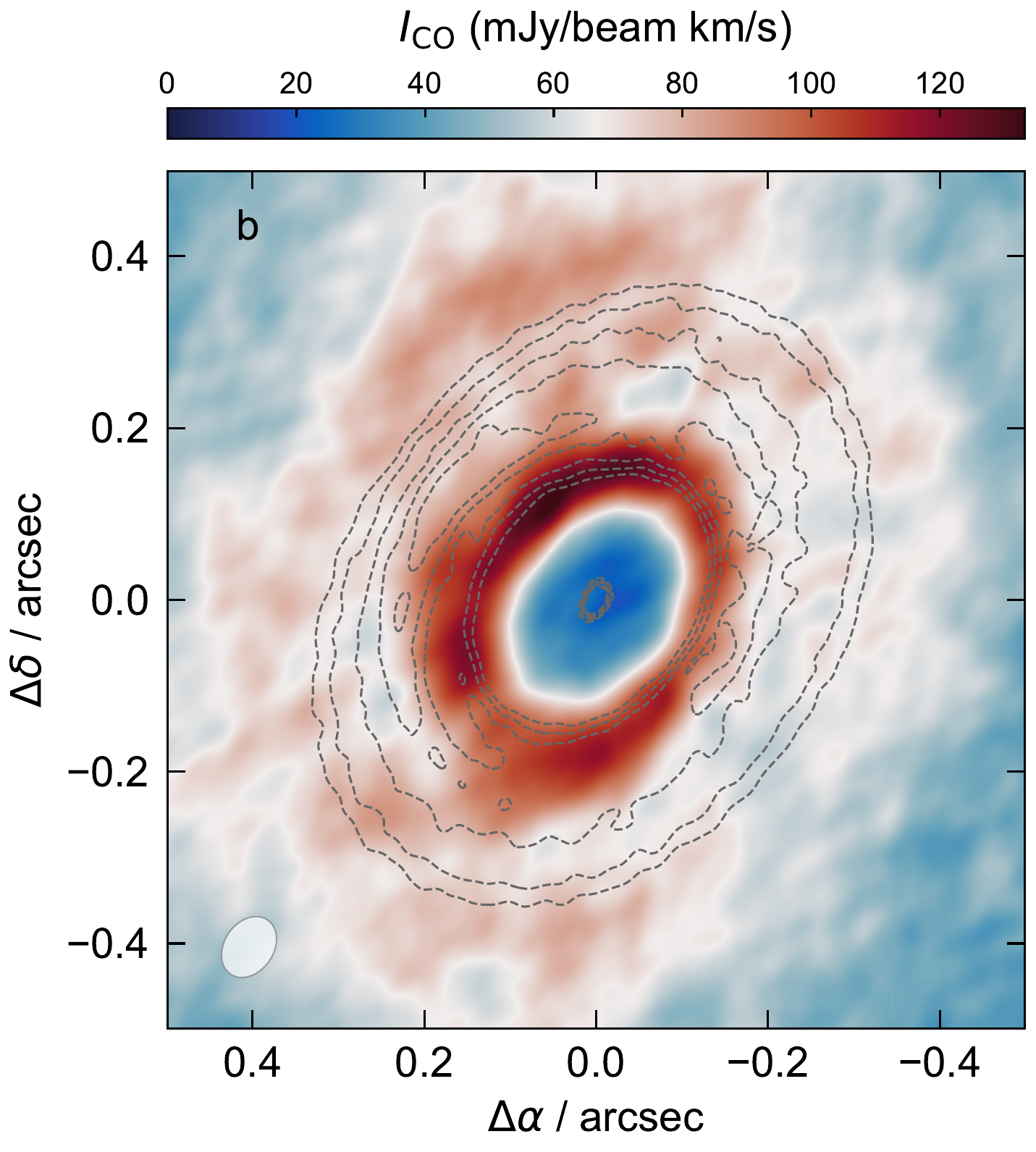}\\
  \includegraphics[width=0.37\textwidth]{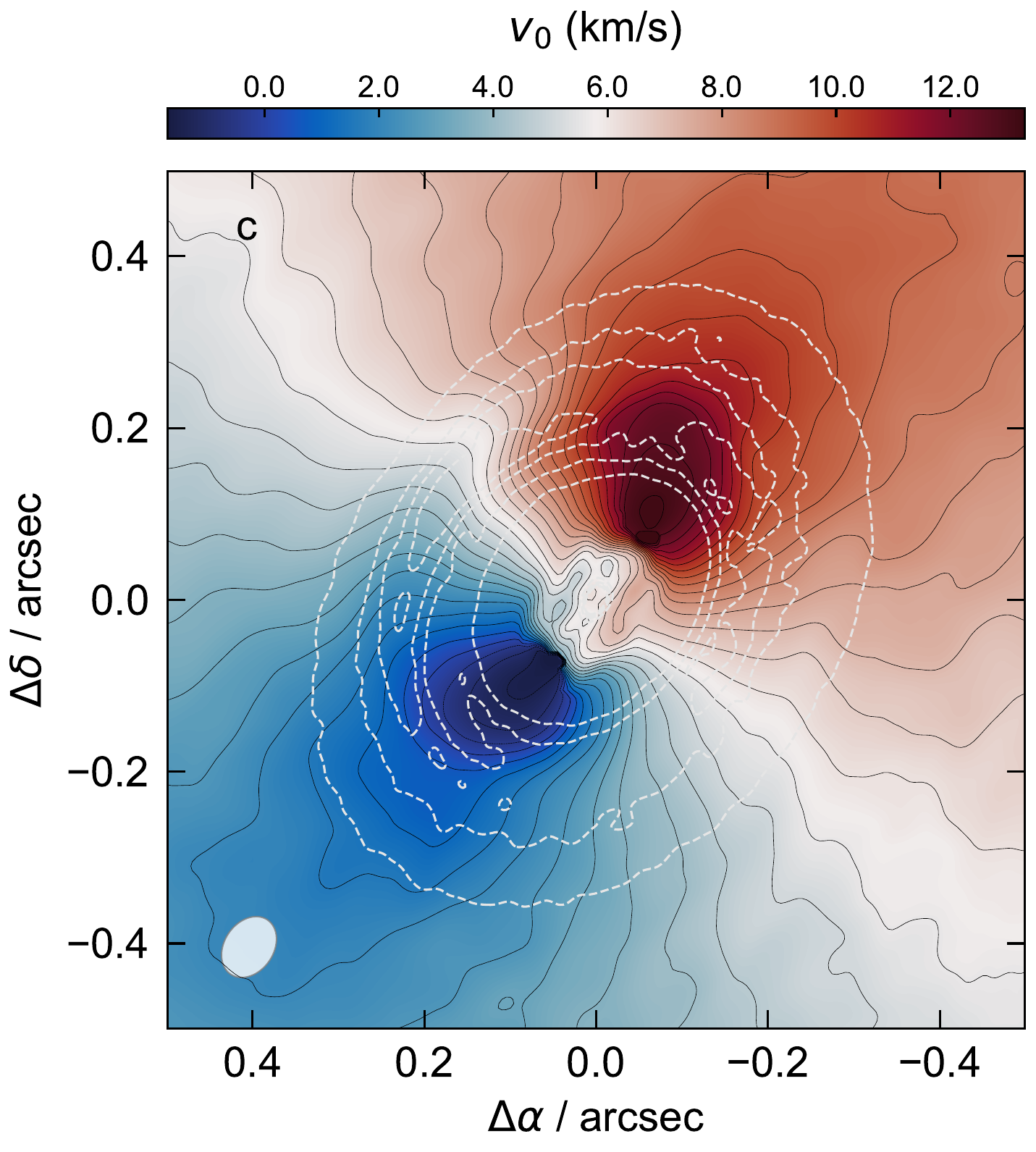}
  \includegraphics[width=0.37\textwidth]{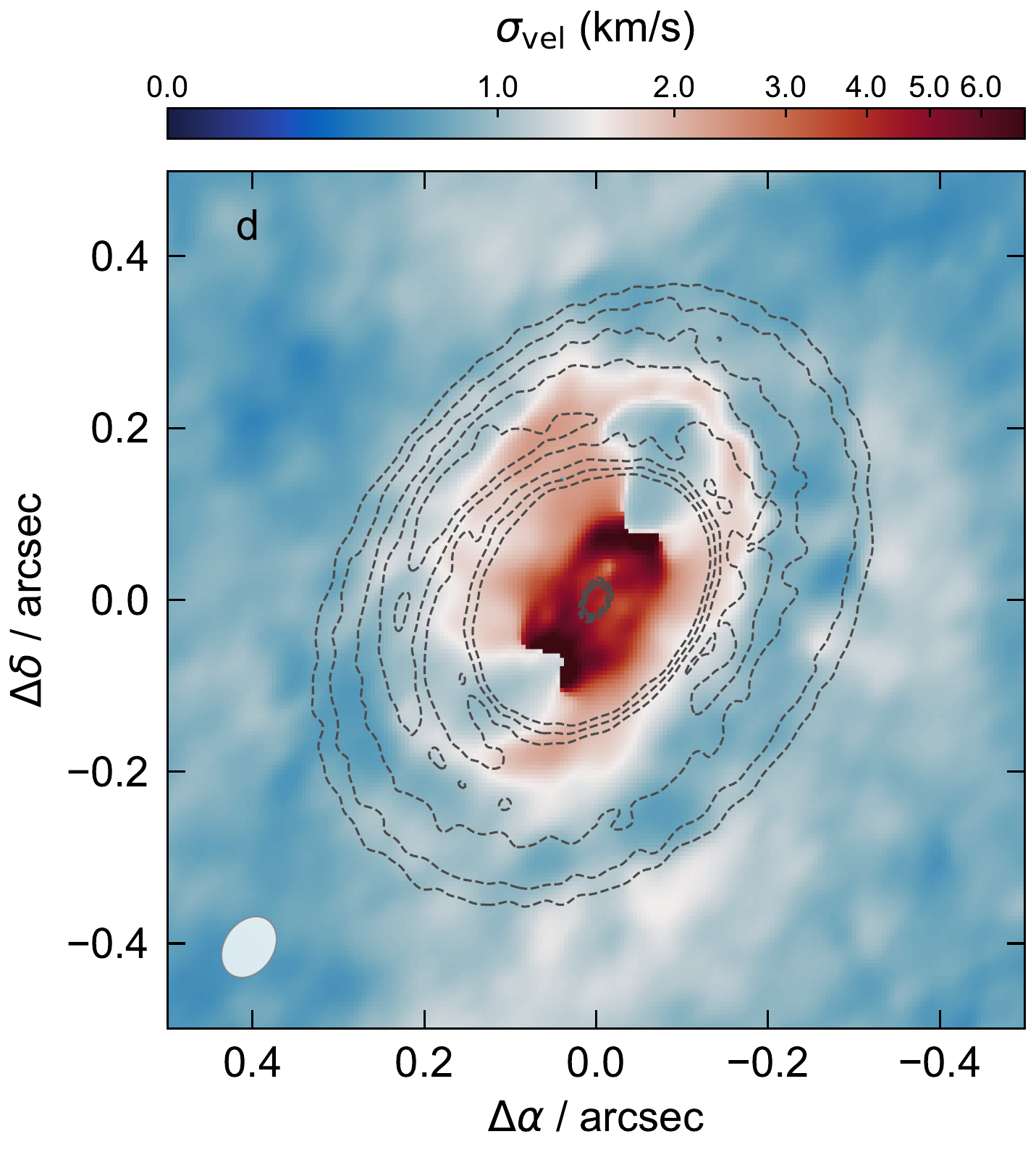}
  \caption{ALMA observations of HD\,100546 acquired in September
    2017. (a) Dust continuum emission map at 1.3 mm
    ($I_{1.3\mathrm{mm}}$). (b) $^{12}$CO intensity map (0th moment,
    $I_{\rm CO}$). (c) Velocity map (1st moment, $v_0$) from
    $^{12}$CO(2--1) emission which shows perturbed non-Keplerian flows
    at the location of the continuum ring. (d) $^{12}$CO velocity
    dispersion map (2nd moment, $\sigma_{\rm vel}$). Contours of
    $I_{1.3\mathrm{mm}}$ are plotted over moment maps with levels
    0.15, 0.27, 0.40, 0.52, and 0.64 mJy~beam$^{-1}$. The lowest
    contour corresponds to 9$\sigma_{\rm MEM}$. All panels display the
    same field-of-view. }
  \label{fig:obs}
\end{figure*}

\section{Results}
\label{sec:results}

\begin{figure*}
  \centering\includegraphics[width=0.9\textwidth]{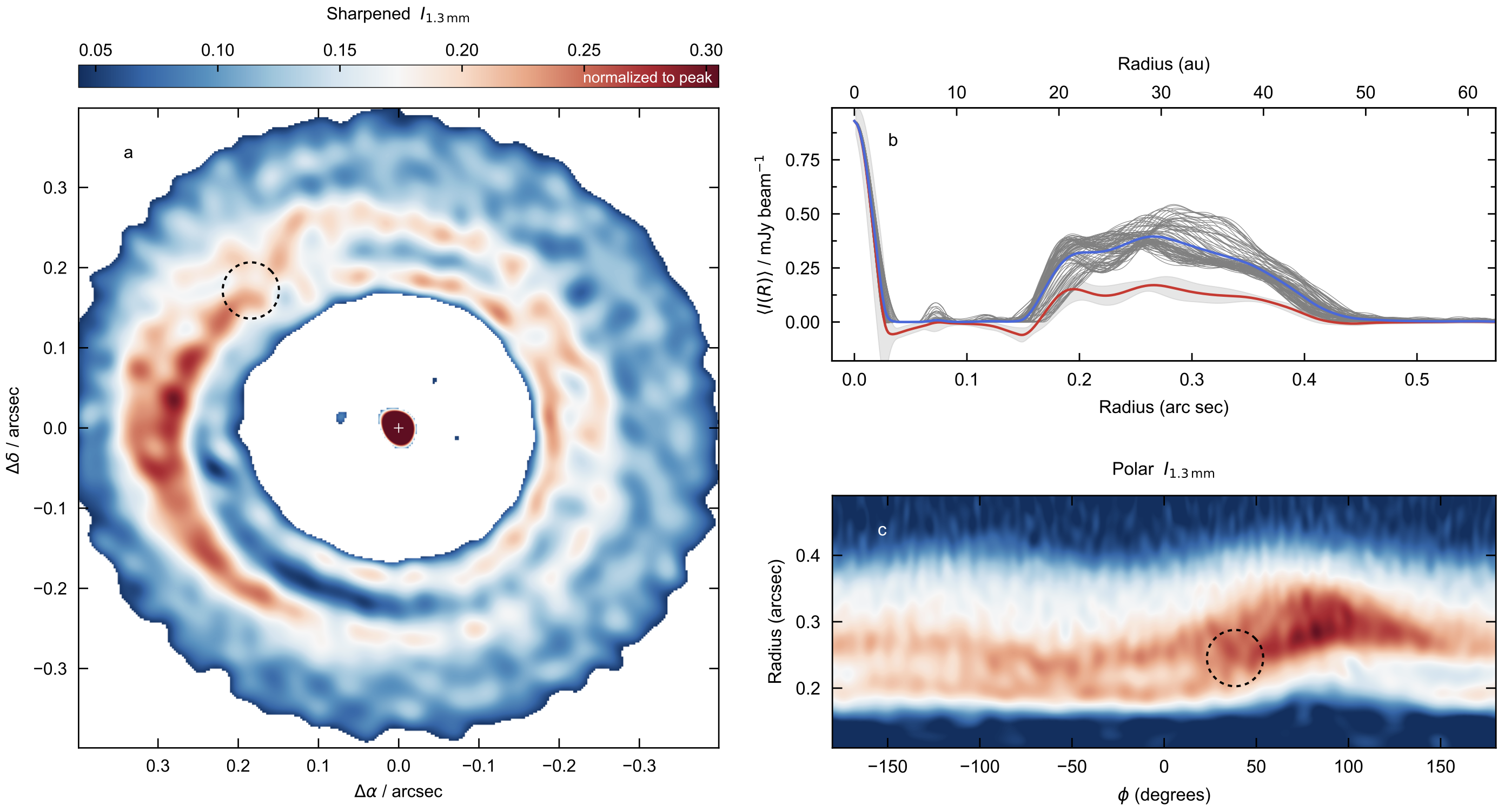}
  \caption{(a) A sharpened version of the 1.3 mm continuum map
    $I_{1.3\mathrm{mm}}$ shown in Figure \ref{fig:obs}a, using an
    unsharp-masking procedure to enhance the fine structure of the
    continuum ring. The image is deprojected to show a face-on
    view. (b) Surface brightness profiles from the original
    $I_{1.3\mathrm{mm}}$ image. The blue solid line corresponds to the
    azimuthally-averaged radial profile, while the family of thin gray
    curves correspond to profiles extracted from 3.6$^\circ$ wedges
    covering the full azimuthal range. The solid red curve shows the
    azimuthally averaged profile of the sharpened imaged shown in
    panel a. The shaded area around the profile corresponds to the
    dispersion around the mean. (c) Polar deprojection of
    $I_{1.3\mathrm{mm}}$, zoomed on the continuum ring. The circle in
    (a) and (c) shows the approximate location of the Doppler flip
    seen in $^{12}$CO gas. }
  \label{fig:sharp}
\end{figure*}

\subsection{A structured continuum ring}
\label{sec:continuum}

The ALMA observations of HD\,100546 are summarized in Figure
\ref{fig:obs}. The first panel (a) shows a bright ring in continuum
emission \citep[the ring has previously been imaged, most recently
  by][in 0.88 mm emission at 50$\times$30~mas
  resolution]{Pineda2019}. In these new images, at 20$\times$13~mas
resolution, the ring displays remarkable radial and azimuthal
substructures. These substructures are further emphasized in an
unsharp-masked version of the continuum image shown in Figure
\ref{fig:sharp}. This sharper image is obtained as the difference
between the original $I_{1.3\mathrm{mm}}$ map and its smoothed
version, after convolution with a circular Gaussian kernel
($\sigma=60$~mas). The procedure is equivalent to removing low-spatial
frequencies and enhances the small-scale features
\citep{Stolker2017}. The sharpened image highlights intriguing breaks
in the arcs conforming the ring, especially at ${\rm PA}=80^\circ$,
where the substructures split into two branching arms. The maze of
ridges may suggest a complex dynamical scenario at play within the
ring.

Figure \ref{fig:sharp} also shows the image transformed into polar
coordinates and its azimuthally-averaged surface brightness radial
profile. The polar deprojection was carried out following the
procedure outlined in \citet{Perez2019}, adopting the disk orientation
derived in \citet{Casassus2019}. The offset between the center of the
ring and the stellar position reveals that the ring is eccentric
\citep[as quantifyied in][]{Pineda2019}.

Continuum emission is detected around the location of the star. The
emission is resolved with a radius of $\sim$1.8\,au (half the FWHM
along its major axis) at 19.7$\times$12.9~mas resolutions. We
interpret this as thermal emission from an asymmetric inner
disk. However, the shape of this central emission could be affected by
the PSF.

Several compact features are present inside the gap (see
Fig.~\ref{fig:sharp}). The brightest of these features is 5$\sigma$,
however, it is most likely an image synthesis artefact due to its
proximity to the sidelobe of the natural-weighted beam. Any other
signal inside the gap are at $\leq$3$\sigma$.

\subsection{Gas kinematics}
\label{sec:kinematics}

\begin{figure*}[t]
  \centering\includegraphics[width=0.95\textwidth]{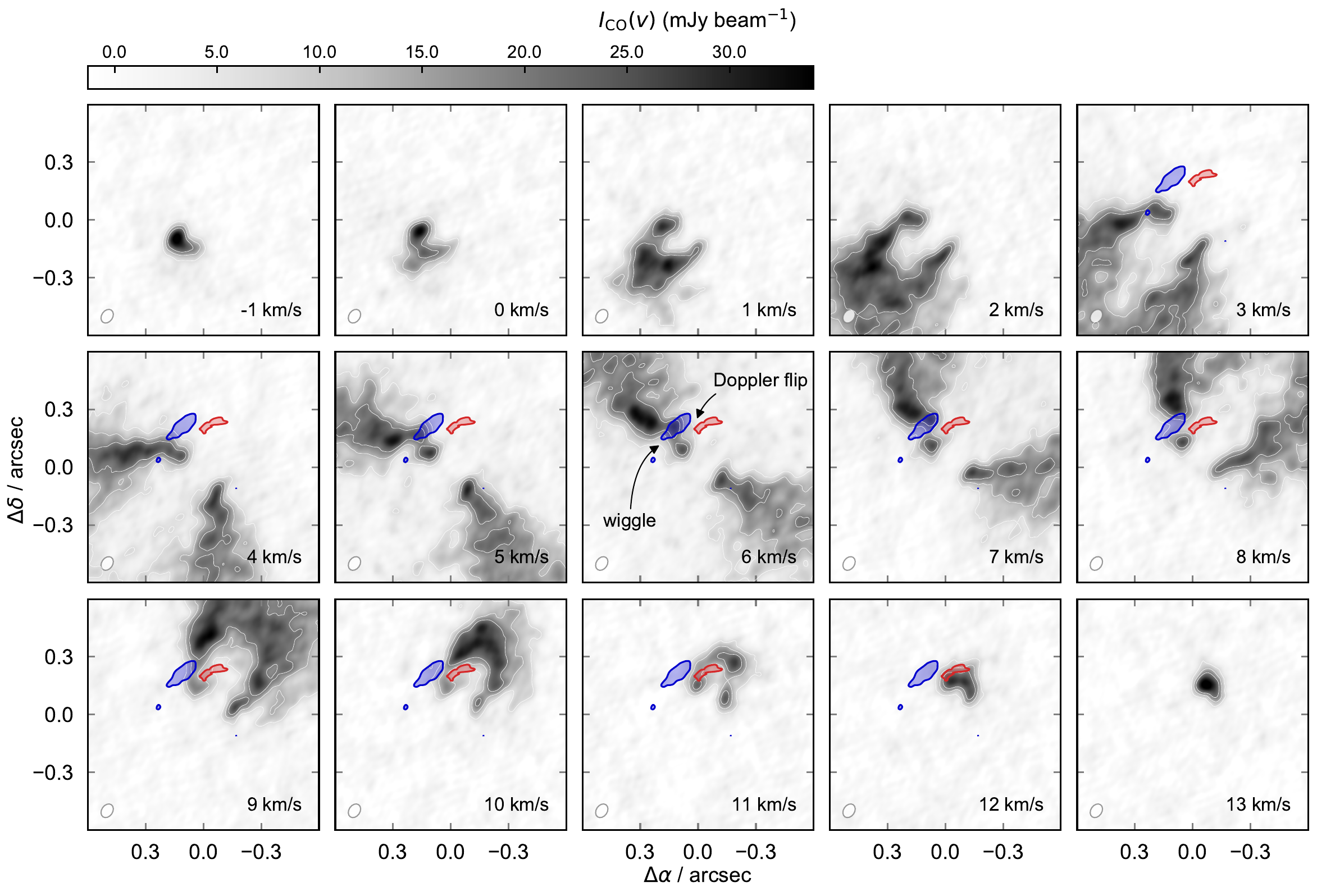}
  \caption{$^{12}$CO emission channel maps of HD\,100546 showing
    several non-Keplerian structures. Most strikingly, it shows
    velocity 'wiggles' (or {\em kinks}) between 0\farcs2 and 0\farcs3
    separation north east from the star. This local non-Keplerian
    feature can be seen in channels between $+$3 and $+$8 km s$^{-1}$
    in velocity. The Doppler flip in the deviation from axially
    symmetric kinematics \citep[presented in][]{Casassus2019} is shown
    as blue and red filled contours. The wiggle and Doppler flip
    labelled in the panel at $+$6 km s$^{-1}$. Each panel shows
    $^{12}$CO emission sampled every 1 km~s$^{-1}$ (although the
    channel width is 0.5 km s$^{-1}$).  Contours correspond to 5, 10,
    and 15 times the channel RMS (1.3 mJy beam$^{-1}$). The
    synthetized beam of the CLEAN reconstruction is shown in the
    bottom left corner. }
  \label{fig:channels}
\end{figure*}

\subsubsection{Kinematic signature over the continuum ring}
\label{sec:wiggle}

The long baseline ALMA observations presented here were part of a
program to reach high enough sensitivities to map the velocity field
of HD\,100546 at fine angular scales. The $^{12}$CO(2--1) moment maps
for intensity (0th moment, $I_{\rm CO}$), velocity field (1st moment,
$v_0$) and velocity dispersion (2nd moment, $\sigma_{\rm vel}$) are
presented in Figure \ref{fig:obs}. Individual channel maps are
presented in Figure \ref{fig:channels}.

The velocity map $v_0$ is shown in Figure \ref{fig:obs}c. At
separations comparable to the continuum ring radius, the iso-velocity
contours in the near side of the disk (south west, bottom right)
display a symmetric pattern with respect to the disk minor
axis. Interestingly, the pattern becomes highly asymmetrical on the
far side (north east, top left), especially near the continuum peak,
approximately 0\farcs24 northeast of the central star. The location of
the wiggles extend over $\sim$90$^\circ$ in azimuth.

As stated earlier, significant deviations from Keplerian motion can be
attributed to planet-disk interactions, via the local flow in a CPD
and the planet-launched spiral wakes. However, we note that other
effects can also produce such wiggles in the iso-velocity
contours. For example, an optically thick continuum, such as the one
in HD\,100546, blocks the CO emission from the back side of the disk,
inducing structure in the gas velocity map. This is indeed the case
for the wiggle-like morphology seen in the south west in the 1st
moment, where the wiggles bear a reflection symmetry with respect to
the disk minor axis.  If the wiggles are due to planet-disk
interaction kinematics, the location of the embedded perturber can be
identified via a local {\em Doppler flip} in molecular line moment 1
maps, after subtraction of the axially symmetric flow
\citep*{Perez2018}, which follows the disk rotation curve. Such sign
reversal in the Keplerian deviation map is indeed associated to the
kinematic signature found between 0\farcs2 and 0\farcs3 in HD\,100546
\citep[this analysis is presented in][]{Casassus2019}.  The magnitude
of the Doppler flip as well as the morphology of the wiggles resemble
that of an embedded massive planet of 5-10~M$_{\rm Jup}$ as shown in
previous hydrodynamical simulations \citep*[see][their
  fig. 3]{Perez2018}.

The wiggles associated with the Doppler flip can be recognized
directly in the iso-velocity channel maps (Figure \ref{fig:channels}),
most clearly between 5.0 and 8.0 km~s$^{-1}$. These wiggles are indeed
easily connected with the `blue' part of the Doppler flip signal. On
the other hand, the red counterpart corresponds to a deficit of signal
between 9 and 11\,km\,s$^{-1}$, replaced by signal at
12\,km\,s$^{-1}$. If the non-Keplerian kinematics are due to a compact
body, the location of the perturber can be pinpointed via the Doppler
flip. As the Doppler flip is located between ridges in the maze of
substructures seen in the continuum ring (see sharpened 1.3 mm map in
Fig.~\ref{fig:sharp}), this would mean the perturber is still embedded
in the dust ring.

\begin{figure*}
  \centering\includegraphics[width=0.95\textwidth]{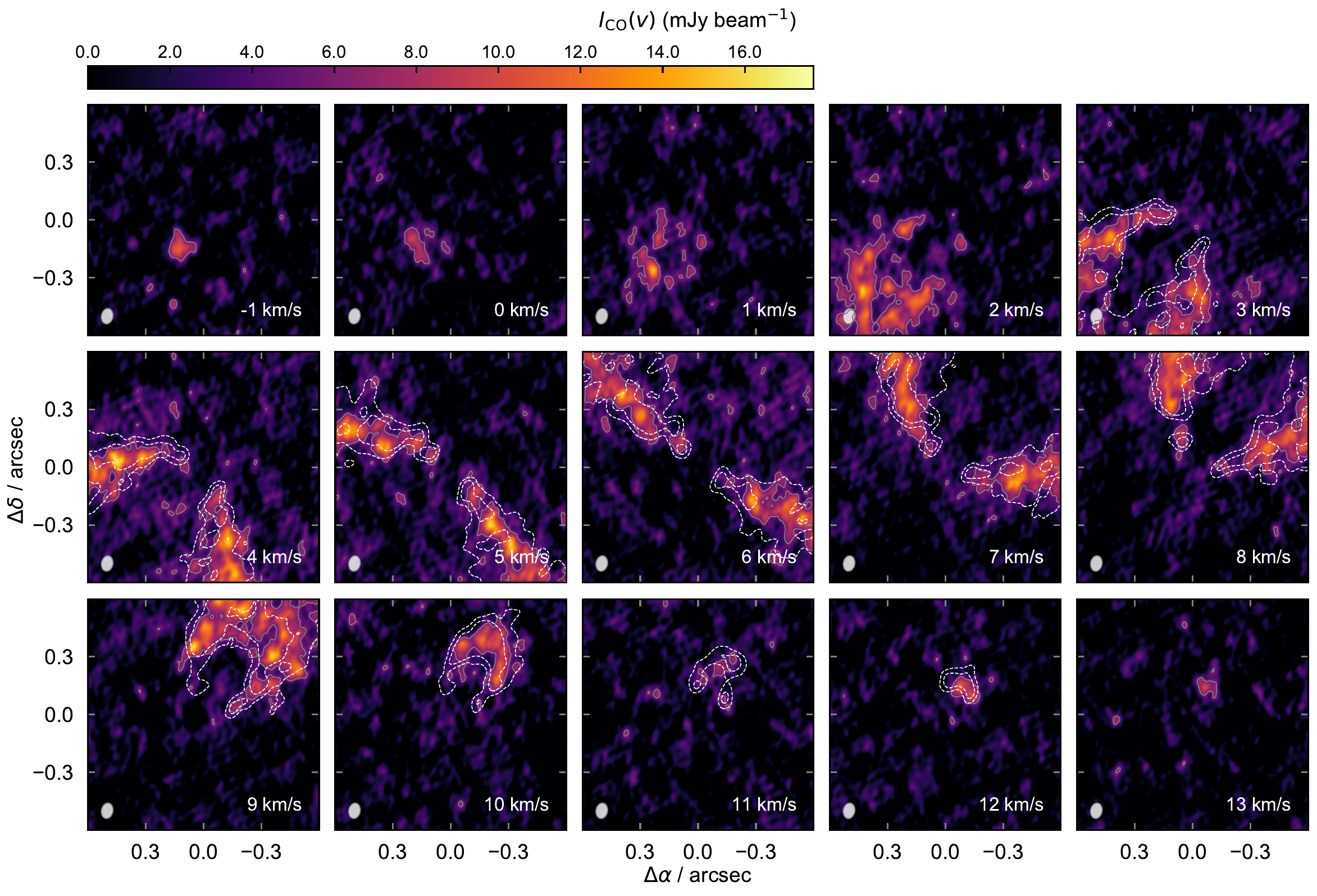}
  \caption{Same as Fig.~\ref{fig:channels} but for $^{13}$CO
    emission. The white dashed contours show the corresponding
    $^{12}$CO emission at 10 and 16 times the $^{12}$CO maps RMS,
    while the gray solid contour shows $^{13}$CO at 3 times the RMS
    level. The 7 km s$^{-1}$ channel most clearly shows that $^{13}$CO
    do not exhibit the same obvious wiggle as in $^{12}$CO. }
  \label{fig:13co}
\end{figure*}

\begin{figure*}
  \centering\includegraphics[width=0.95\textwidth]{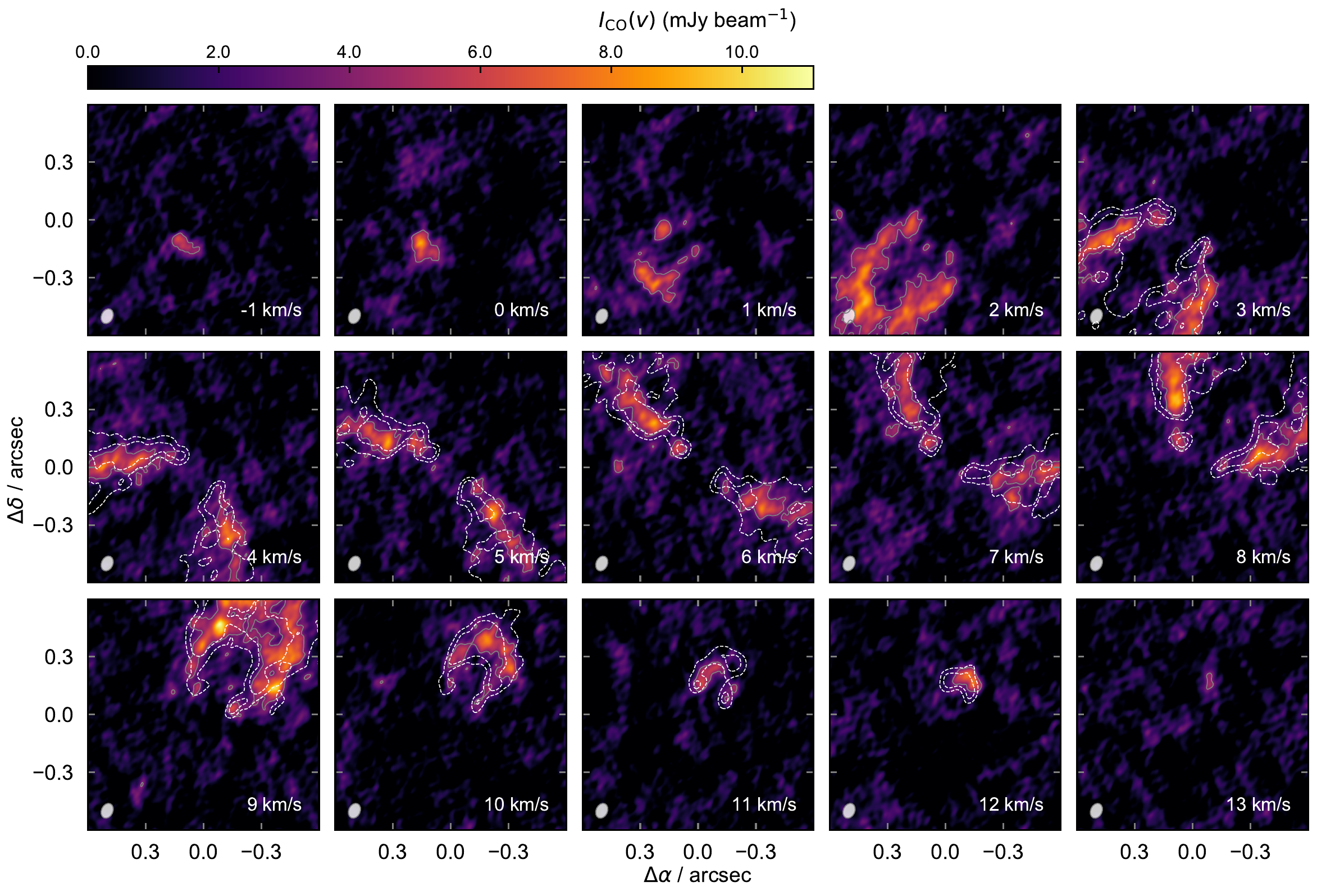}
  \caption{Same as Fig.~\ref{fig:13co} but for C$^{18}$O emission.  }
  \label{fig:c18o}
\end{figure*}

\subsubsection{Vertical structure of the wiggle via isotopologue emission}

Fig.~\ref{fig:13co} shows that the $^{13}$CO channel maps do not
exhibit any obvious wiggles, as seen in $^{12}$CO. This is most easy
to notice in the 7 km~s$^{-1}$ panel where the $^{13}$CO emission does
not follow the morphology of the clear wiggle seen in $^{12}$CO (shown
as white dashed contours). At 6 km~s$^{-1}$ there is only a hint for a
kinematic deviation in $^{13}$CO, while the wiggle is notorious in
$^{12}$CO. The same can be said for the C$^{18}$O emission shown in
Fig.~\ref{fig:c18o}. These rarer CO isotopologues have lower optical
depths and thus trace deeper layers in the disk, suggesting that the
kinematic signature weakens at lower heights. Interestingly, the three
isotopologues show a decrement in emission at the location where the
Doppler flip transitions from blue to red.

In an idealized disk as used in hydrodynamic simulations the vertical
velocities must cancel at the midplane due to mirror symmetry. On the
other hand, observational arguments suggest that the bulk of the
Doppler flip signal is either due to radial or vertical flows
\citep{Casassus2019}. Thus, the qualitative absence of kinematic
wiggles in the rarer isotopologues is consistent with them being
mostly due to vertical motions in the disk surface.

\subsubsection{Non Keplerian kinematics within the gap}
\label{sec:gas}

The $^{12}$CO intensity $I_{\rm CO}$ shows an inner depletion of gas
emissions within the continuum gap. Although the gap is entirely
cleared in the dust, it appears filled of CO gas. Within the gap,
diffuse CO emission averages 30 mJy~beam$^{-1}$ km~s$^{-1}$ with a
scatter of 6 mJy~beam$^{-1}$ km~s$^{-1}$. This is not inconsistent
with a planetary origin for the gap clearing as residual gas is
expected to remain and be detectable in optically thick tracers such
as CO \citep[e.g.][]{Ober2015, Facchini2018}.

The gas flow within the dust gap is highly perturbed at radii
$<$60~mas (see Figure \ref{sec:obs}c). Within this radius, the gas
rotation pattern twists its position angle by almost 90$^\circ$, more
than could be explained by stellocentric accretion, even at free-fall
rates \citep[e.g.][]{Casassus2015}. These deviations are manifested
along all position angles, hinting at large scale kinematic effect
such as strong radial inflows and a warp, or reflecting planet-disk
interaction kinematics, which can also explain deviations covering
wide azimuthal ranges with an accreting giant, as shown in
\citet*{Perez2018}. If a perturber in an inclined orbit is present
within the gap, it could drive warping of the inner regions, as in the
cavity of HD\,142527 \citep{Casassus2015, Price2018}. HD\,142527
indeeds harbours a $\sim$0.2$M_\odot$ star which has been detected
with SAM observations \citep{Biller2014} and H$\alpha$ high-contrast
imaging \citep{Close2014}.

Several companion candidates have been reported in HD\,100546.
\citet{Quanz2013a} present candidate ``b'' at 0\farcs5, while
\citet{Currie2015} identified a weakly-polarized disk feature or
candidate ``c''. As mentioned in the introduction, these
direcly-imaged IR detections are being debated \citep{Thalmann2016,
  Follette2017, Hord2017, Sissa2018}. On the other hand, spectroscopic
monitoring of rovibrational CO emission shows variability which is
consistent with a companion orbiting near the edge of the gas cavity
\citep{Brittain2014, Brittain2019}. Although this companion would
orbit farther away from the star ($\sim$0\farcs1) than the deviation
in kinematics seen within the cavity ($\sim$60~mas), it may well be
connected to the origins of gap and the non-Keplerian low-$J$
$^{12}$CO emission.

The velocity dispersion map $\sigma_{\rm vel}$ shows arm-like
structures on top of the continuum ring (Fig.~\ref{fig:obs}d). This is
also expected if the gap is being carved by a giant planet, via the
observable kinematics of meridional circulations
\citep{Dong2019}. Within the gap, the line broadens abruptly at
$\sim$50~mas at the same location that the velocity field twists its
PA.  A comparison with hydrodynamic predictions is being developed and
will be presented in a future publication.

\subsection{Could the gap be opened by a stellar object?}
\label{sec:sam}

\begin{figure*}
  \centering \includegraphics[height=0.35\textwidth]{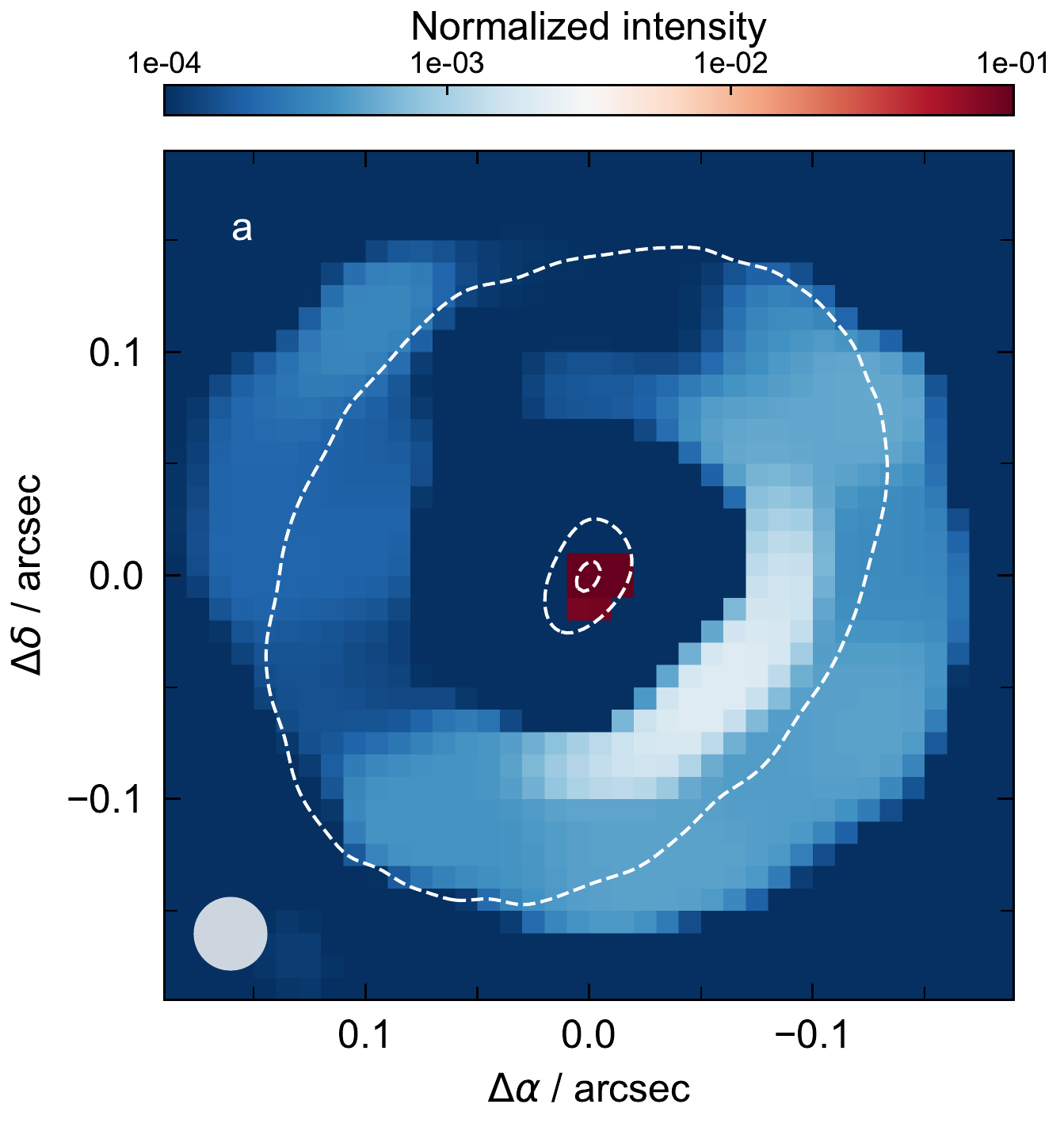}
  \includegraphics[height=0.35\textwidth]{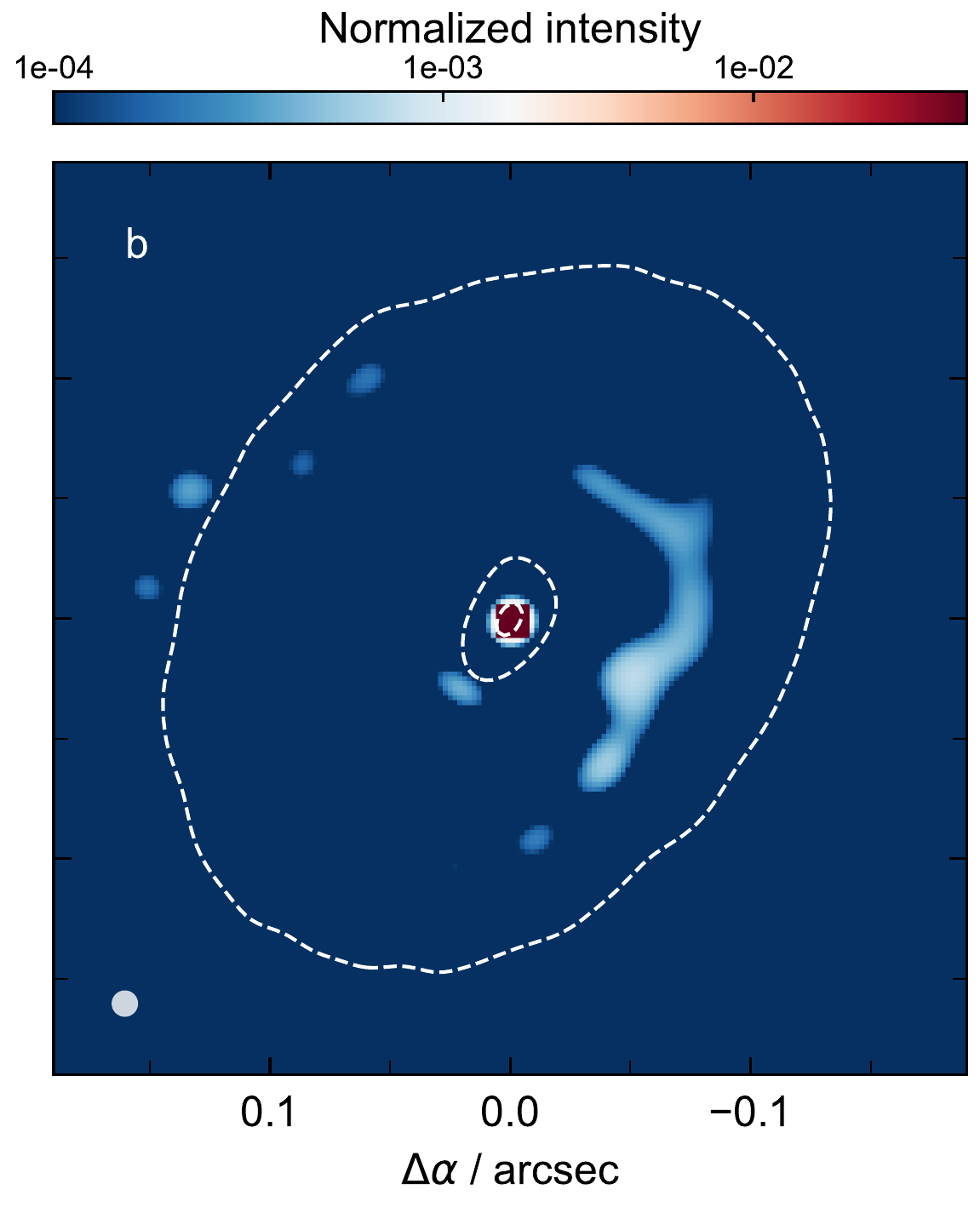}
  \hfill\includegraphics[height=0.24\textwidth]{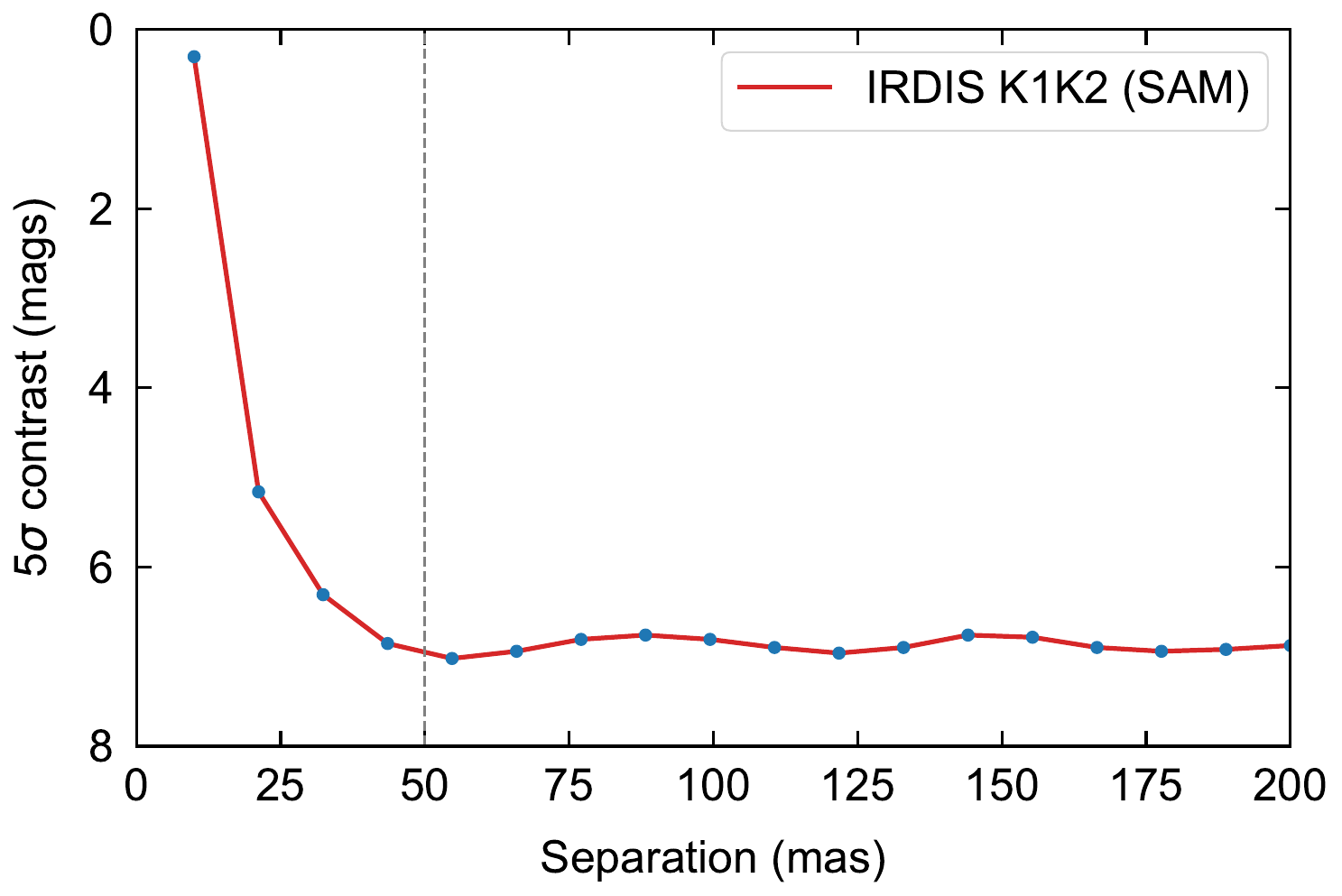}
  \caption{SPHERE SAM observations of HD\,100546. The IRDIS (a) and
    IFS (b) images were produced with the MIRA package, applied to the
    closure phases and visibilities, with a hyperbolic regularization
    term. The images show extended scattered light and an empty gap
    free of stellar companions. The dashed contour shows the
    9$\sigma_{\rm MEM}$ level from the $I_{1.3\mathrm{mm}}$ continuum
    image. (c) IRDIS $5\sigma$ contrast curve for SAM observations in
    K1K2 band. A contrast of $\sim$7 mags is achived between 50 and
    200 mas. }
  \label{fig:sam}
\end{figure*}

The SAM observations (Figure \ref{fig:sam}) reveal extended signal
(probably scattered light) from the disk, and a protoplanetary gap of
smaller radius than the one in mm continuum (here shown in dashed
contours). Rather, the scattered light signal seems to delineate a
region within the gap which matches the radius at which the gaseous
velocity field becomes highly perturbed.  The IRDIS image recovers
emission from every PA, while IFS only yields signal from the maximum
forward scattering angle (towards the south west).  The gap is empty
at the achieved contrast level.

At the separation of $<$50~mas, the contrast in magnitude is 6.9
(Figure~\ref{fig:sam}c). To convert this 5$\sigma$ contrast in upper
limits on the mass of putative companions, we used the BT-Settl models
by \citep{2012RSPTA.370.2765A}, adapted for SPHERE filters. This limit
corresponds to a mass of 33 M$_{\mathrm{Jup}}$ if we assume the
youngest age of 4 Myr for the star, or 71 M$_{\mathrm{Jup}}$ for the
eldest of 12 Myr. We can exclude the presence of stellar companions
around HD\,100546.

\section{Summary and conclusions}
\label{sec:discussion}

In this Letter we presented high resolution 1.3 mm observations of
HD\,100546 with ALMA. The continuum reveals an optically thick and
rather wide ring, extending from 15 to 45~au in radius. The
unprecedented angular resolutions, in this source, revealed a network
of ridges, hitherto unseen in any such ring.

The velocity field of this protoplanetary disk shows strong signatures
of non Keplerian flows, most remarkably in the form of strong wiggles
in the channel maps, from 0\farcs2 to 0\farcs3 north east from the
star, and extending over almost 90\,deg in azimuth. The spiral wakes
of an accreting giant planet could affect the line-of-sight velocities
in a way that resembles the observed wiggles.

The $^{12}$CO channel maps show that the wiggles trace the
blue-shifted emission associated to a Doppler flip in the deviation
from axially symmetric kinematics \citep[presented
  in][]{Casassus2019}. This sign reversal is expected from planet-disk
interaction simulations. The red-shifted Doppler flip emission is
difficult to identify in the global kinematics by inspecting the
channel maps alone. Thus, observationally and without recourse to
hydrodynamic simulations, the location of the presumed perturber
cannot easily be pin-pointed by the wiggles but rather by the Doppler
flip.

The qualitative absence of kinematic wiggles in the rarer
isotopologues is consistent with them being weaker at lower disk
heights, and suggests that most of the perturbed flow is associated to
vertical motions in the disk surface. Our observations also show
enhanced dispertion in velocity at the stellocentric radius of the
Doppler flip, which could reflect meridional flows in the context of a
planet-disk interaction \citep{Teague2019}.

It is remarkable that the kinematic signature coincides with bright
continuum emission. This is counterintuitive as an accreting planet is
expected to open a deep gap in the dust density field, which is why we
did not attempt to produce a matching hydrodynamical simulation with a
single giant \citep[as in][]{Pinte2018}. Such a configuration may
perhaps be reproduced with a closer-in giant \citep[such as a that
  proposed by][]{Brittain2019}, which sustains the ring structure
while a rapidly accreting giant planet is embedded within the ring.

\acknowledgments

We thank the anonymous referee for a very constructive report and for
spotting an important flaw in the original manuscript. Financial
support was provided by the government of Chile grants Millennium
Scientific Initiative RC130007, CONICYT-Gemini 32130007, and
CONICYT-FONDECYT grant numbers 1171624, 1171246 and 1191934. S.P
acknowledges support from the Joint Committee of ESO and the
Government of Chile. A.Z. acknowledges support from CONICYT PAI 2017
folio PAI77170087. This project has received funding from the European
Union's Horizon 2020 research and innovation programme under grant
agreement No 748544 (PBLL). This paper makes use of the following ALMA
data: ADS/JAO.ALMA\#2016.1.00344.S. ALMA is a partnership of ESO
(representing its member states), NSF (USA) and NINS (Japan), together
with NRC (Canada), MOST and ASIAA (Taiwan), and KASI (Republic of
Korea), in cooperation with the Republic of Chile. The Joint ALMA
Observatory is operated by ESO, AUI/NRAO and NAOJ. The National Radio
Astronomy Observatory is a facility of the National Science Foundation
operated under cooperative agreement by Associated Universities, Inc.

\vspace{5mm}
\facilities{ALMA Observatory}
\facilities{VLT/SPHERE}

\software{GPUVMEM \citep{Carcamo2018}, CASA \citep{McMullin2007}, MIRA
  \citep{Thibaut2008}.}



\bibliography{hd100546}

\end{document}